\documentclass[aps,prd,preprint]{revtex4-1}
\usepackage{amsmath}
\usepackage{amssymb}
\usepackage{revsymb}

\begin{document}

\title{Very Special Relativity and Lorentz Violating Theories}

\author{Jorge Alfaro}
\affiliation{Facultad de F\'\i sica, Pontificia Universidad
Cat\'olica de Chile,
 Casilla 306, Santiago 22, Chile}
 \email[E-mail: ]{jalfaro@puc.cl}
\author{Victor O. Rivelles}
\affiliation{Instituto de F\'\i sica, Universidade de S\~ao Paulo,
C. Postal 66318,  S\~ao Paulo, SP, Brazil}
\email[E-mail: ]{rivelles@fma.if.usp.br }



\begin{abstract}Very special relativity (VSR) keeps the main features of special relativity but breaks rotational invariance. We will show how VSR like terms which depend on a fixed null vector can be generated systematically.  We start with a formulation for a spinning particle which incorporates VSR. We then use this formulation to derive the VSR modifications to the Maxwell equations. Next we consider VSR corrections to Thomas precession. We start with the coupling of the spinning particle to the electromagnetic field adding a gyromagnetic factor which gives rise to a magnetic moment. We then  propose a spin vector in terms of the spinning particle variables and show that it obeys the BMT equation. All this is generalized to the VSR context and we find the VSR contributions to the BMT equation. 
\end{abstract}



\maketitle

\section{Introduction}

The standard model of particle physics is a well established and experimentally confirmed theory but it needs to be extended in order to incorporate some known phenomena like, for instance,  neutrino masses and dark matter. Possibly the standard model is the low energy limit of a larger theory which hopefully includes gravity. Since the present experimental data are not enough to point out how to extend the standard model we must seek for small deviations of it which could be detected at low energies. One possibility is that fields from a more complete theory couple to the standard model fields like constant background fields causing deviations of Lorentz symmetry \cite{Colladay:1998fq}. This is a very active line of investigation with many theoretical results awaiting for experimental confirmation (for a review see \cite{Liberati:2013xla}). Usually such proposals have as a consequence that the dispersion relation for light is modified. A more conservative alternative would keep the essential features of special relativity, like the constancy of the velocity of light, but leave aside rotation symmetry for instance. This can be achieved by taking subgroups of the Lorentz group which preserve the constancy of the velocity of light. Such subgroups were identified and used to built what is called very special relativity (VSR) \cite{Cohen:2006ky}. One of its mains feature is that the inclusion of $P$, $T$ or $CP$ symmetries enlarges VSR to the full Lorentz group so that VSR is only relevant in theories where one of the discrete symmetries is broken. Two subgroups of the Lorentz group, $SIM(2)$ and $HOM(2)$, have the property of rescaling a fixed null vector $n^\mu$. Then terms containing ratios of contractions of $n^\mu$ with other kinematic vectors will be invariant under transformations of these subgroups. A proposal to generate mass for neutrinos along these lines was presented in \cite{Cohen:2006ir} where an equation for a left-handed fermion incorporating VSR was given 
\begin{equation} \label{1.1}
	\left( \slash\!\!\! p  - \frac{1}{2} m^2 \frac{\slash\!\!\!n}{n^\mu p_\mu} \right) \psi_L = 0, 
\end{equation}
where $m$ sets the VSR scale. When the equation of motion is squared we find that it describes a free fermion of mass $m$. The price to be paid is the presence of non-local operators as well as the lack of rotational symmetry. In this way it is possible to save some of the important effects of special relativity and consider possible violations of space isotropy. Several aspects of VSR have been considered, like the inclusion of supersymmetry \cite{Cohen:2006sc,Vohanka:2011aa}, curved spaces \cite{Gibbons:2007iu,Muck:2008bd}, noncommutativity \cite{SheikhJabbari:2008nc,Das:2010cn}, dark matter \cite{Ahluwalia:2010zn} and also in cosmology \cite{Chang:2013xwa}.


We can take this specific realization of VSR and consider the addition of interaction terms in the context of the usual Lorentz violating theories.  We can regard the inclusion of operators containing a constant and null vector $n^\mu$ as determining a preferred direction in space. It breaks Lorentz symmetry to $ISO(2)$ but allowing a scale transformation on $n^\mu$ the symmetry can be enlarged to $SIM(2)$. So the inclusion of terms containing ratios of $n^\mu$ contracted with other kinematic vectors will lead to the consideration of VSR like terms as that in (\ref{1.1}). When a Lorentz invariant  action is extended by the addition of Lorentz violating operators the coefficients of such operators are in general arbitrary and unrelated to each other. In this paper we will show that Lorentz violating terms, like the one present in (\ref{1.1}), can be derived in a systematic way.  To do that we start in Section (\ref{s1}) with the model of a massive spinning particle describing a free fermion. It is characterized by its worldline reparametrization and worldline supersymmetry. To give rise to a VSR term similar to that in (\ref{1.1}) the supersymmetry constraint is modified. This is the only point where a Lorentz violating term is added by hand. In Section (\ref{s3}) we consider a spinning particle with ${\cal N}=2$ supersymmetries which describes an abelian gauge field. In order to derive the Lorentz violating terms contributing to the Maxwell equations we consider the modified supersymmetry constraint from the previous section. We find a massive photon in agreement with the VSR construction done in \cite{Cheon:2009zx}. This approach provides a systematical way of generating Lorentz violating terms like the one in (\ref{1.1}). 

We then apply this approach to interacting theories.  To be concrete we consider the relativistic equation describing Thomas precession also known as BMT equation \cite{Bargmann:1959gz,Jackson}. It describes the dynamics of an axial 4-vector $S^\mu$, associated to the spin of the electron, in the presence of an uniform electromagnetic field and from it it is possible to derive the precession angular velocity of the spin in the electron rest frame. In order to apply our formalism we have to construct $S^\mu$ in terms of the spinning particle variables and this is done in the next section where we consider the coupling of the usual spinning particle to the Maxwell field and define a Grassmannian spin vector $S^\mu$ for the spinning particle. We then show how it naturally leads to the BMT equation. Then in  Section \ref{s5} we use the VSR spinning particle obtained in Section \ref{s1} to derive corrections to the BMT equation. We find that the spin vector $S^\mu$ must have additional terms depending on $n^\mu$. We work in the limit where $m$ is much smaller than the electron mass and find that many new terms contribute  to the BMT equation. As expected the coefficients of the Lorentz breaking terms are not arbitrary, the only arbitrariness being the value of $m$. Finally, in Section \ref{s6} we present some conclusions.

\section{VSR Spinning Particle} \label{s1}

The spinning particle \cite{Berezin:1976eg} provides a particle description for a Dirac field in the same way as the relativistic particle is associated to the Klein-Gordon field. Besides the particle coordinates $X^\mu(\tau)$ and its momentum $P^\mu(\tau)$ we also need Grassmannian coordinates $\Psi^\mu(\tau)$ and $\Psi_5(\tau)$ satisfying Poisson brackets
\begin{equation}
\{ X^\mu, P_\nu\} = \delta^\mu_\nu, \qquad \{ \Psi^\mu, \Psi^\nu \} = \frac{i}{2}\eta^{\mu\nu}, \qquad \{\Psi_5, \Psi_5\} = -\frac{i}{2}.
\end{equation}
We assume the existence of a first class constraint ${\cal S}$ which generates worldline supersymmetry
\begin{equation}\label{2.2}
	{\cal S} = P_\mu \Psi^\mu - M \Psi_5.
\end{equation}
We then find that the Poisson bracket algebra closes on the Hamiltonian constraint
\begin{equation}
	\{ {\cal S}, {\cal S}  \} = i {\cal H}, \qquad {\cal H} =  \frac{1}{2} (P^2 - M^2).
\end{equation}

The quantization is performed by promoting the Poisson brackets to commutators or anticommutators
\begin{equation}
	[ X^\mu, P_\nu ] = -i \delta^\mu_\nu, \qquad \{\Psi^\mu, \Psi^\nu \} = \frac{1}{2} \eta^{\mu\nu}, \qquad 
	\{ \Psi_5, \Psi_5 \} = - \frac{1}{2},
\end{equation}
so that $P_\mu = i \partial_\mu$ and the Grassmannian variables are proportional to the Dirac gamma matrices $\Psi^\mu = \frac{1}{2} \gamma^\mu \gamma_5, \Psi_5 = \frac{1}{2} \gamma_5$. Then the physical states $\varphi(x)$ must satisfy the supersymmetry constraint ${\cal S} \varphi(x) = 0$ and we get the massive Dirac equation. 

In VSR the massive Dirac equation for a fermion is modified to \cite{Cohen:2006sc}
\begin{equation}\label{2.5}
	\left( i \slash\!\!\!\partial + \frac{i}{2} m^2 \frac{\slash\!\!\!n}{n \partial} - M \right) \varphi(x) = 0,
\end{equation}
where $n^2=0$, $m$ is the VSR mass scale and $n \partial = n^\mu \partial_\mu$. We will use the notation that for two vectors $A^\mu$ and $B^\mu$, $AB = A^\mu B_\mu$. This strongly suggests that we modify the supersymmetry constraint (\ref{2.2}) to 
\begin{equation}\label{2.6}
	{\cal S} = P \Psi - \frac{1}{2} m^2 \frac{\Psi n}{P n} - M \Psi_5,
\end{equation}
so that the Poisson bracket algebra still closes on the Hamiltonian constraint which is modified to
\begin{equation}
	{\cal H} = \frac{1}{2} ( P^2 - m^2 - M^2).
\end{equation}
Then the effect of VSR like term in the supersymmetry constraint is just a shift in the squared particle mass. Notice also that the supersymmetry constraint is still well behaved with respect to VSR transformations since $n^\mu$ appears on the numerator and denominator of the new term. 

The action for the VSR spinning particle has the standard form 
\begin{equation}\label{2.8}
	S = \int \, d\tau \, ( P \dot{X} - i \Psi \dot{\Psi} + i \Psi_5 \dot{\Psi_5} + e {\cal H} + i \chi {\cal S} ),
\end{equation}
where $e(\tau)$ and $\chi(\tau)$ are Lagrange multipliers. From the constraints we can derive the infinitesimal worldline supersymmetry  transformations
\begin{eqnarray}
\delta X^\mu &=& -i \epsilon \left( \Psi^\mu + \frac{1}{2} m^2 \frac{\Psi n}{(P n)^2} n^\mu \right), \\
\delta P^\mu &=& 0, \\
\delta \Psi^\mu &=& - \frac{1}{2} \epsilon \left( P^\mu - \frac{1}{2} m^2 \frac{n^\mu}{P n} \right), \\
\delta \Psi_5 &=& - \frac{1}{2} M \epsilon, \\
\delta e &=& -i \epsilon \chi, \\
\delta \chi &=& \dot{\epsilon},
\end{eqnarray}
and worldline reparametrization transformations 
\begin{eqnarray}
\delta X^\mu &=&  \xi P^\mu, \\
\delta e &=& \dot{\xi} \\
\delta P^\mu &=& \delta \Psi^\mu = \delta \Psi_5 =  \delta \chi = 0,
\end{eqnarray}
where $\epsilon$ is a Grassmannian supersymmetry parameter and $\xi$ is the reparametrization parameter. The action is invariant under these transformations up to a total derivative term. Upon quantization the wave function has to satisfy the supersymmetry constraint (\ref{2.6}) and we get the VSR Dirac equation (\ref{2.5}). 


\section{Maxwell Equations in VSR}\label{s3}

Since the Dirac equation is modified in VSR the same must happen to the Maxwell equations. To show this we can use the spinning particle with extended supersymmetry. The general case was treated in \cite{Howe:1988ft} where it was shown that a spinning particle with ${\cal N}$ supersymmetries describes massless field equation for particles with spin ${\cal N}/2$. A path integral analysis was performed in \cite{Pierri:1990rp}. Here we will consider the case ${\cal N}=2$ in the context of VSR.

We consider two Grassmannian variables $\Psi^\mu_i$, $i=1,2$, and the following constraints
\begin{eqnarray}
    {\cal H} &=& \frac{1}{2} (P^2 - m^2) \\
	{\cal S}_i &=& P \Psi_i - \frac{1}{2} m^2 \frac{\Psi_i n}{Pn},\\
	{\phi}_{ij} &=& \Psi_i \Psi_j.
\end{eqnarray}
The constraint ${ \phi}_{ij}$ generates $SO(2)$ rotations so we have two supersymmetries. The constraint algebra is
\begin{eqnarray}
	\{ {\cal S}_i, {\cal S}_j \} &=& \delta_{ij} {\cal H}, \\
	\{ \phi_{ij}, {\cal S}_k \} &=& {\cal S}_i \delta_{jk} - {\cal S}_j \delta_{ik}, \\
	\{ \phi_{ij}, \phi_{kl} \} &=& \delta_{ik} \phi_{jl} - \delta_{il} \phi_{jk} + \delta_{jk} \phi_{il} - \delta_{jl} \phi_{ik}. 
\end{eqnarray}

The physical states $\varphi$ must satisfy all constraints. The anticommutation relations
\begin{equation}
	\{ \Psi^\mu_i, \Psi^\nu_j \} = \eta^{\mu\nu} \delta_{ij},
\end{equation}
can be realized in terms of gamma matrices as \cite{Howe:1988ft}
\begin{equation}
	\Psi^\mu_1 = \gamma^\mu \otimes 1, \qquad \Psi^\mu_2 = \gamma_5 \otimes \gamma^\mu.
\end{equation}
This means that the physical states are bispinors $\varphi_{\alpha\beta}$. Then the $SO(2)$ constraint implies that $\varphi_{\alpha\beta} = (\sigma^{\mu\nu} C)_{\alpha\beta} F_{\mu\nu}(x)$, where $C$ is the charge conjugation matrix and $(\sigma^{\mu\nu} C)_{\alpha\beta}$ is  symmetric in the spinor indices. 

The constraint ${\cal S}_i$ implies that 
\begin{equation}
	\displaystyle{\not} \partial_\alpha^\beta \varphi_{\beta\gamma} + \frac{1}{2} m^2 \frac{\displaystyle{\not} n_\alpha^\beta}{n \partial} \varphi_{\beta\gamma} = 0.
\end{equation}
We can rewrite this equation for $F_{\mu\nu}$ getting
\begin{equation}
	\left( \partial_\mu F_{\nu\lambda} + \frac{1}{2} m^2 \frac{n_\mu}{n \partial} F_{\nu\lambda} \right) (\gamma^\mu \sigma^{\nu\lambda})_\alpha^\beta = 0.
\end{equation}
Since $\gamma^\mu \sigma^{\nu\lambda}$ is proportional to $\epsilon^{\mu\nu\lambda\rho} \gamma_\rho \gamma_5$ and $\eta^{\mu[\nu} \gamma^{\lambda]}$ we can take the trace to get
\begin{equation}\label{29}
	\partial_{[\mu} F_{\nu\lambda]} + \frac{1}{2} \frac{m^2}{n\partial} n_{[\mu} F_{\nu\lambda]} = 0,
\end{equation}
while multiplying by $\gamma_5$ and taking the trace we get
\begin{equation}\label{30}
	\partial^\mu F_{\mu\nu} + \frac{1}{2} \frac{m^2}{n\partial} n^\mu F_{\mu\nu} = 0.
\end{equation}
In special relativity when $m^2=0$ we recover the Bianchi identities and the Maxwell equations. In VSR they are modified. They also imply that
\begin{equation}
\square F_{\mu\nu} + m^2 F_{\mu\nu} = 0, 
\end{equation}
showing that $F_{\mu\nu}$ has mass $m$. 

We can try to solve the VSR  Bianchi identities (\ref{29}) and remarkably there is a solution
\begin{equation}
	F_{\mu\nu} = D_{[\mu} A_{\nu]}, \qquad D_\mu = \partial_\mu + \frac{1}{2} \frac{m^2}{n \partial} n_\mu.
\end{equation}
Notice that $D_\mu$ has an abelian algebra but it does not satisfy the Leibniz rule. 
Notice also that  $F_{\mu\nu}$ is not invariant under the usual gauge transformation but it is invariant under
\begin{equation}\label{33}
	\delta A_\mu = D_\mu \Lambda.
\end{equation}
Then the VSR Maxwell equations (\ref{30}) can be written as
\begin{equation}
	D^\mu F_{\mu\nu} = 0,
\end{equation}
and we have a massive field described by a field equation with a modified gauge invariance (\ref{33}). Our results agree with those found in \cite{Cheon:2009zx}.

The non-Abelian extension of VSR gauge fields was done in \cite{Alfaro:2013uva}. It was found that since all gauge fields in a given multiplet acquire a common mass $m$ it can not be used as a replacement for the Higgs mechanism.

\section{Coupling the Spinning Particle to the Maxwell Field and the  BMT Equation}\label{s2}

In this section we show how to derive the BMT equation in special relativity using the spinning particle variables. Firstly we couple the spinning particle to a background electromagnetic field $A_\mu$ using the minimal substitution  $P^\mu \rightarrow P^\mu - q A^\mu$ in the supersymmetry constraint (\ref{2.2}). To introduce the gyromagnetic factor $g$ we consider the proposal  for a spinning particle with ``anomalous" magnetic moment \cite{Barducci:1982yw}. There is a more detailed treatment in \cite{Gitman:1992an} where the expressions are more explicit. Besides the minimal substitution we also have to replace $M \rightarrow M + 2i \mu F\Psi\Psi$, in the supersymmetry constraint, where the magnetic moment is
\begin{equation}
	\mu = \frac{q}{2M} ( \frac{g}{2} - 1),
\end{equation}
so we get 
\begin{equation}\label{4.2}
	{\cal S} = \Psi (P - q A) - (M + 2i\mu F\Psi\Psi) \Psi_5. 
\end{equation}
The notation $FAB = F_{\mu\nu} A^\mu B^\nu$ is used throughout the rest of the paper. The Hamiltonian constraint is now
\begin{equation}
	{\cal H} = \frac{1}{2} [ (P - qA)^2 - M^2 ] - i( q + 2 \mu M)F\Psi\Psi - 4i\mu F(P- qA)\Psi \Psi_5 + 2\mu^2 (F\Psi\Psi)^2.
\end{equation}
It can also be checked that $\{ {\cal H}, {\cal S} \}=0$. 

The action has the same form as before (\ref{2.8}) and the equations of motion are
\begin{eqnarray}
P^\mu &=& qA^\mu - \frac{1}{e} \dot{X}^\mu + 4i\mu F^{\mu\nu}\Psi_\nu \Psi_5 -  \frac{i}{e} \chi\Psi^\mu, \label{38} \\
\dot{P}^\mu &=& e [ -q (P^\nu - qA^\nu) \partial^\mu A_\nu - i (q + 2\mu M) \partial^\mu F\Psi\Psi - 4i\mu (\partial^\mu F)(P-qA)\Psi \Psi_5  \nonumber \\ 
&-& 4i\mu q (\partial^\mu A^\nu) F_{\rho\nu}\Psi^\rho\Psi_5 + 4 \mu^2 F\Psi\Psi \partial^\mu F\Psi\Psi ] - iq\chi \Psi\partial^\mu A, \\
\dot{\Psi}^\mu &=& e[ -(q + 2\mu M) F^{\mu\nu}\Psi_\nu + 2\mu F^{\mu\nu}(P-qA)_\nu \Psi_5 - 4 i \mu^2 F\Psi\Psi F^{\mu\nu} \Psi_\nu ] \nonumber \\
&-& \frac{1}{2} \chi [ (P-qA)^\mu - 8i\mu F^{\mu\nu}\Psi_\nu \Psi_5], \\
\dot{\Psi}_5 &=&  -2\mu e F(P-qA)\Psi - \frac{1}{2} \chi (M + 2i \mu F\Psi\Psi), 
\end{eqnarray}
plus the constraints. We now choose the gauge $e = -1/M$ and  $\chi =0$. Since we are interested only in weak and uniform background fields we can linearize the above equations. Also eliminating the momentum we find
\begin{eqnarray}
\ddot{X}^\mu &=& \frac{q}{M} F^{\mu\nu} \dot{X}_\nu, \label{4.8} \\
\dot{\Psi}^\mu &=& (\frac{q}{M} + 2\mu) F^{\mu\nu}\Psi_\nu - 2\mu F^{\mu\nu}\dot{X}_\nu\Psi_5\label{4.9}, \\
\dot{\Psi}_5 &=& 2\mu F\dot{X}\Psi, \\
\dot{X}\Psi &+& 2i\frac{\mu }{M} F\Psi\Psi \Psi_5 - \Psi_5 = 0, \label{45} \\
\dot{X}^2 &-& 1 - 2 \frac{i}{M} ( \frac{q}{M} + 2\mu) F\Psi\Psi = 0. \label{4.12}
\end{eqnarray}

The next step is to define an axial spin vector $S^\mu(\tau)$ which generalizes the rest frame spin of the electron. Requiring that its time component vanishes in the rest frame it must satisfy $\dot{X} S = 0$. There are several proposals to describe the relativistic spin through some particle model (see for instance \cite{Deriglazov:2011gy}). Here we assume that $S^\mu$ is a pseudo-vector even in the Grassmannian variables and the natural choice is 
\begin{equation}
	S^\mu = \epsilon^{\mu\nu\rho\sigma} \dot{X}_\nu \Psi_\rho \Psi_\sigma. 
\end{equation}
When computing $\dot{S}^\mu$ we have to rewrite all terms quadratic in $\Psi$ in terms of $S$. To do that we use the identity
\begin{equation}\label{4.14}
	\Psi^\mu \Psi^\nu = \frac{1}{2\dot{X}^2} \epsilon^{\mu\nu\rho\sigma} \dot{X}_\rho S_\sigma - \dot{X}^{[\mu} \Psi^{\nu]} \frac{\dot{X}\Psi}{\dot{X}^2},
\end{equation}
where $A_{[\mu} B_{\nu]} = A_\mu B_\nu - A_\nu B_\mu$ with no factor of 1/2. 
We also have to use the field equations (\ref{4.9}-\ref{4.12}) noting that $\dot{\Psi}^\mu, \dot{\Psi}_5, \dot{X}^2 - 1$ and $\dot{X}\Psi - \Psi_5$ are all of ${\cal O}(F)$. The calculation is lengthy and tedious. There are several terms proportional to $\epsilon^{\mu\nu\rho\sigma}\dot{X}_\nu \Psi_\sigma \Psi_5$ which can not be rewritten in terms of $S^\mu$ but cancel against each other. At the end the result is 
\begin{equation}
\dot{S}^\mu = \frac{qg}{2M} \left( F^{\mu\nu}S_\nu +  \dot{X}^\mu FS\dot{X} \right) - \dot{X}^\mu S\ddot{X}.
\end{equation}
Using now the equation of motion (\ref{4.8}) we get the BMT equation 
\begin{equation}
	\dot{S}^\mu = \frac{q}{M} \left( \frac{g}{2} F^{\mu\nu}S_\nu + (\frac{g}{2} - 1) \dot{X}^\mu FS\dot{X} \right).
\end{equation}
Having obtained the BMT equation from the spinning particle the next step is to generalize it to VSR since we already know the supersymmetry constraint (\ref{2.6}). 

\section{Coupling the VSR Spinning Particle}\label{s5}

We go along the same lines as in the previous section. The simplest choice for the supersymmetry constraint which reduces to (\ref{4.2}) and (\ref{2.6}) is 
\begin{equation}
	{\cal S} = \Psi (P - q A) - (M + 2i\mu F\Psi\Psi) \Psi_5 - \frac{1}{2} m^2 \frac{\Psi n}{(P-qA)n}. 
\end{equation}
The Poisson bracket algebra of two supersymmetries constraints closes on 
\begin{eqnarray} \label{52}
	{\cal H} &= \frac{1}{2} [ (P - qA)^2 - m^2 - M^2 ] - i( q + 2 \mu M)F\Psi\Psi - 4i\mu F(P- qA)\Psi \Psi_5 + 2\mu^2 (F\Psi\Psi)^2 \nonumber \\
	 &+ iq m^2 \Psi n \frac{ F\Psi n}{[(P-qA)n]^2} - 2i\mu m^2 \frac{F\Psi n}{(P-qA)n} \Psi_5 + 2\mu m^2 \Psi n \frac{ n^\rho \partial_\rho F\Psi\Psi}{[(P-qA)n]^2} \Psi_5.
\end{eqnarray}
Again it is possible to show that its Poisson bracket with ${\cal S}$ vanishes. 

When deriving the equations of motion we have to deal with $(P-qA)n$ in several denominators. To do that we take the field equation obtained by varying $P^\mu$ 
\begin{eqnarray}
	(P-qA)^\mu &=& - \frac{1}{e} \dot{X}^\mu + 4i\mu F^{\mu\nu}\Psi_\nu\Psi_5 + 2m^2 \left( iq \Psi n \frac{F\Psi n}{[(P-qA)n]^3} - i\mu \frac{F\Psi n \Psi_5}{[(P-qA)n]^2}  \right. \nonumber  \\
  &+& \left. 2\mu \Psi n \frac{n^\rho \partial_\rho F\Psi\Psi}{[(P-qA)n]^3} \Psi_5 \right) n^\mu - \frac{i}{e} \chi \left( \Psi^\mu + \frac{1}{2} m^2 \frac{\Psi n}{[(P-qA)n]^2} n^\mu \right),
\end{eqnarray}
and contract it with $n^\mu$ so that 
\begin{equation}
	(P-qA)n = - \frac{\dot{X} n}{e} \left( 1 + 4i\mu e \frac{F\Psi n}{\dot{X}n} \Psi_5 + i \chi \frac{ \Psi n}{\dot{X} n} \right).  
\end{equation}
We can now invert this equation taking into account that we have Grassmannian variables inside the parenthesis 
\begin{equation}
	\frac{1}{(P-qA)n} = - \frac{e}{\dot{X}n} \left( 1 - 4i\mu e \frac{F\Psi n}{\dot{X}n} \Psi_5 - i \chi \frac{\Psi n}{\dot{X} n} + 8\mu e \frac{F\Psi n\Psi_5}{(\dot{X}n)^2} \chi \Psi n \right).
\end{equation}

Since the particle has mass $\sqrt{m^2 + M^2}$ we now choose the gauge $e = -1/\sqrt{m^2 + M^2}$ and $\chi =0$. The linearized equations of motion become
\begin{eqnarray}
\ddot{X}^\mu &=&\frac{q}{\sqrt{m^2 + M^2}} F^{\mu\nu} \dot{X}_\nu, \label{5.6}\\
\dot{\Psi}^\mu &=& \frac{q + 2\mu M}{\sqrt{m^2 + M^2}}  F^{\mu\nu}\Psi_\nu - 2\mu F^{\mu\nu}\dot{X}_\nu\Psi_5 - \frac{q}{2} \frac{m^2}{(m^2 + M^2)^{3/2}} \frac{1}{(\dot{X} n)^2} \left( F\Psi n \,\, n^\mu - \Psi n \,\, F^{\mu\nu}n_\nu \right) \nonumber \\
&+& \mu \frac{m^2}{m^2 + M^2}  \frac{F^{\mu\nu} n_\nu}{\dot{X}n} \Psi_5, \\
\dot{\Psi}_5 &=& 2\mu F\dot{X}\Psi + 2 \mu \frac{m^2}{m^2 + M^2} \frac{F\Psi n}{\dot{X}n}, \\
\dot{X}\Psi &+& 2 \frac{\mu }{\sqrt{m^2 + M^2}} F\Psi\Psi \Psi_5 - \frac{M}{\sqrt{m^2 + M^2}}\Psi_5  - \frac{1}{2} \frac{m^2}{m^2 + M^2} \frac{\Psi n}{\dot{X} n} = 0, \\
\dot{X}^2 &-& 1 - \frac{2i}{m^2 + M^2} ( q + 2\mu M) F\Psi\Psi + 6iq \frac{m^2}{m^2 + M^2} \frac{\Psi n}{(\dot{X} n)^2} F\Psi n = 0.
\end{eqnarray}
Notice that the Lorentz force law in VSR (\ref{5.6}) keeps the same form as in special relativity and does not depend on $n^\mu$. Just the mass has changed to include the VSR mass scale $m$.


The next step is to compute $\dot{S}^\mu$. Besides the identity (\ref{4.14}) we will need another identity obtained from the former one by contracting it with $n^\mu$. After using the equations of motion it reads
\begin{equation}
	\Psi_\mu \Psi n = \frac{2(m^2+M^2)}{m^2+2M^2} \left( \frac{1}{2(\dot{X})^2} \epsilon_{\mu\nu\rho\sigma} n^\nu \dot{X}^\rho S^\sigma + \frac{M}{\sqrt{m^2+M^2}} \dot{X}n \Psi_\mu \Psi_5 \right) + \dots,
\end{equation}
where $\dots$ are terms proportional to $\dot{X}^\mu$ which do not contribute to the relevant calculations. It is then found that the cancellation among the $\epsilon^{\mu\nu\rho\sigma}\dot{X}_\nu \Psi_\sigma \Psi_5$ terms no longer occurs and $\dot{S}$ can not be written in terms $S$. The only way out is to modify the definition of $S^\mu$. 

In fact having a new vector $n^\mu$ allows the construction of other vectors out of a bilinear in the Grassmannians. For instance 
\begin{equation}
\tilde{S}^\mu = \frac{1}{\dot{X} n} \epsilon^{\mu\nu\rho\sigma} \dot{X}_\nu n_\rho \Psi_\sigma \Psi_5
\end{equation}
satisfies $\dot{X}\tilde{S}=0$ so it is a candidate. Another possibility is $\epsilon^{\mu\nu\rho\sigma} n_\nu \Psi_\rho \Psi_\sigma$ which does not vanish when contracted with $\dot{X}$ but with $n$. It is possible to multiply it by a projector so that it vanishes when contracted with $\dot{X}$ 
\begin{equation}
	\hat{S}^\mu = \frac{1}{\dot{X} n} \epsilon^{\mu\nu\rho\sigma} n_\nu \Psi_\rho \Psi_\sigma - \frac{\dot{X}^\mu}{\dot{X}^2} \frac{1}{\dot{X}n} \epsilon^{\lambda\nu\rho\sigma} \dot{X}_\lambda n_\nu \Psi_\rho \Psi_\sigma.
\end{equation}
It turns out that $\hat{S}$ can be written as a combination of $S$ and $\tilde{S}$ as
\begin{equation}
	\hat{S}^\mu =  2 \frac{m^2+M^2}{m^2+2M^2} S^\mu + 4 \frac{M\sqrt{m^2+M^2}}{m^2+2M^2} \tilde{S}^\mu -\frac{m^2}{m^2+2M^2} \frac{S n}{\dot{X} n} \dot{X}^\mu + \frac{m^2}{m^2+2M^2} \frac{S n}{(\dot{X} n)^2} n^\mu.
\end{equation}
We then found that the only combination $S$ and $\tilde{S}$ that gets rid of the $\epsilon^{\mu\nu\rho\sigma}\dot{X}_\nu \Psi_\sigma \Psi_5$ terms mentioned above is given by 
\begin{equation} \label{65}
	S_T^\mu = S^\mu - \frac{m^2}{M\sqrt{m^2+M^2}} \tilde{S}^\mu = \epsilon^{\mu\nu\rho\sigma} \dot{X}_\nu \left( \Psi_\rho\Psi_\sigma - \frac{m^2}{M\sqrt{m^2+M^2}} \frac{1}{\dot{X}n} n_\rho\Psi_\sigma\Psi_5 \right).
\end{equation}
The factor $- \frac{m^2}{M\sqrt{m^2+M^2}}$ is essential for the cancellation. To show that we need further identities like 
\begin{eqnarray}
\epsilon_{\mu\nu\rho\sigma} \dot{X}^\rho \tilde{S}^\sigma &=& \frac{ \dot{X}_{[\mu} n_{\nu]} }{\dot{X} n} \dot{X}\Psi \Psi_5 - \dot{X}_{[\mu} \Psi_{\nu]} \Psi_5 + \dot{X}^2 \frac{ n_{[\mu} \Psi_{\nu]} }{\dot{X}n} \Psi_5, \\
	\epsilon_{\mu\nu\rho\sigma} n^\rho \tilde{S}^\sigma &=& \frac{ \dot{X}_{[\mu} n_{\nu]} }{\dot{X}n}  \Psi n \Psi_5 + n_{[\mu} \Psi_{\nu]} \Psi_5.
\end{eqnarray}

Since the VSR scale is very small we can consider only the case $m^2<<M^2$ and keep terms up to order $m^2/M^2$. In this case we get after a long calculation 
\begin{eqnarray} \label{68}
	\dot{S}^\mu_T &=& \frac{1}{M} \left(1-\frac{1}{2}\frac{m^2}{M^2} \right) (q+2\mu M) F^{\mu\nu}S_{T\nu} + 2\mu \left(\dot{X}^\mu - \frac{1}{2}\frac{m^2}{M^2}\frac{n^\mu}{\dot{X}n}\right) FS_T\dot{X} \nonumber \\  
	&+& \mu \frac{m^2}{M^2} F^{\mu\nu}\dot{X}_\nu \frac{S_Tn}{\dot{X}n} + \frac{q}{2} \frac{m^2}{M^3} F^{\mu\nu} n_\nu \frac{S_Tn}{(\dot{X}n)^2} + \frac{q}{2} \frac{m^2}{M^3} \left(\dot{X}^\mu - \frac{n^\mu}{\dot{X}n} \right) \frac{FS_Tn}{\dot{X}n} \nonumber \\
	&-& \frac{q}{2} \frac{m^2}{M^3} \dot{X}^\mu F\dot{X}n \frac{S_Tn}{(\dot{X}n)^2}.
\end{eqnarray}
This is the generalization of the BMT equation to VSR. As anticipated there are several terms that can be built out of $n^\mu$ but all the coefficients are determined. A consistency check is to notice that 
\begin{equation}
	\dot{X} \dot{S}_T = \frac{q}{\sqrt{m^2+M^2}} F\dot{X}S_T.
\end{equation}

We now have to go to the electron rest frame by a Lorentz boost and $n^\mu$ has to be transformed as well. It can be checked that $S_T \dot{S}_{T}=0$ so that in the rest frame $\vec{S}_T \cdot \dot{\vec{S}}_T =0$ which means that the spin is precessing in that frame. This has been explicitly verified. Then it is possible to compute the VSR corrections to Thomas precession and also the VSR contribution to the anomalous magnetic moment of the electron. 

An alternative way to derive the extension of the BMT equation to VSR is by making use of the distribution function for the spinning particle. In order to relate quantities depending on the Grassmannian variables with observable quantities it is usual to define a distribution function in phase space  \cite{Berezin:1976eg}. The distribution $\rho(\Psi,\Psi_5,t)$ must satisfy a Liouville equation 
\begin{equation} \label{70}
	\frac{\partial \rho}{\partial t} + \{ H, \rho \} = 0,
\end{equation}
and must be normalized to one
\begin{equation}
	\int d\Psi_5 d\Psi_3 d\Psi_2 d\Psi_1 d\Psi_0 \,\, \rho(\Psi,\Psi_5) = 1.
\end{equation}
It is used to define the averaged value of a dynamical variable $F(\Psi,\Psi_5)$ as 
\begin{equation}
<F> \, = \int d\Psi_5 d\Psi_3 d\Psi_2 d\Psi_1 d\Psi_0 \,\, F(\Psi,\Psi_5) \, \rho(\Psi,\Psi_5). 
\end{equation}
In this context the Grassmannian variables are regarded as independent variables so that the supersymmetry constraint ${\cal S}$ is used only at the end of all calculations. In the relativistic case $P \Psi$ and $\Psi_5$ are gauge degrees of freedom so that the distribution function is given by \cite{Berezin:1976eg}
\begin{equation} \label{73}
	\rho = \frac{1}{2} \left( v(t) \Psi + \frac{1}{3} \epsilon^{\mu\nu\rho\sigma} \frac{P_\mu}{M} \Psi_\nu \Psi_\rho \Psi_\sigma \right) \delta\left(\frac{P \Psi}{M}\right) \delta\left(\Psi_5\right), 
\end{equation}
where $v(t)$ satisfies $P v= 0$ and the coefficient $1/3$ is required by normalization.  The distribution function is  defined for the free spinning particle and interactions are introduced in the Hamiltonian in (\ref{70}). Then $P^\mu = M \dot{X}^\mu$ and (\ref{73}) reduces to 
\begin{equation}\label{74}
	\rho = \frac{1}{2} \left( v(t) \Psi + \frac{1}{3} \epsilon^{\mu\nu\rho\sigma} \dot{X}_\mu \Psi_\nu \Psi_\rho \Psi_\sigma \right) \dot{X} \Psi \, \Psi_5,
\end{equation}
with $\dot{X} v = 0$. 

If we consider the VSR contributions to the spinning particle as part of the interactions then our distribution function is (\ref{74}) and  we can use it to compute the averaged value of $S^\mu_T$  (\ref{65}).  We then find that  $<S^\mu_T> = v^\mu$. We now use (\ref{70}) to find the equation satisfied by $<S^\mu_T>$. To this end we get the  Hamiltonian $H$ from (\ref{2.8}) as $H = - e {\cal H} = {\cal H}/\sqrt{m^2+M^2}$ with ${\cal H}$ given by (\ref{52}). Computing the Poisson brackets and using the constraints (\ref{45}) and (\ref{4.12}) and eliminating the momentum using (\ref{38}) we find that the equation satisfied by $<S^\mu_T>$ when $m^2 << M^2$  is exactly (\ref{68}). This provides a powerful check that our extension of the BMT equation to VSR is in the right direction. 

Alternatively we could had started with a distribution function for the VSR spinning particle which already takes into account the VSR effects as described in Section \ref{s1}. Now the gauge degrees of freedom are $\Pi \Psi$ and $\Psi_5$, where 
\begin{equation}
	 \Pi^\mu = P^\mu - \frac{1}{2} m^2 \frac{n^\mu}{Pn}
\end{equation}
so that the distribution function is 
\begin{equation}
	\rho = \frac{1}{2} \left( v \Psi + \frac{1}{3}  \epsilon^{\mu\nu\rho\sigma} \frac{\Pi_\mu}{M} \Psi_\nu \Psi_\rho \Psi_\sigma \right) \frac{\Pi \Psi}{M} \Psi_5,
\end{equation}
with $\Pi v =0$. We again replace the momenta getting
\begin{equation}
	\Pi^\mu = \sqrt{m^2 + M^2} \dot{X}^\mu - \frac{1}{2} \frac{m^2}{\sqrt{m^2 + M^2}} \frac{n^\mu}{\dot{X} n}. 
\end{equation}
Now the averaged value of $S^\mu_T$ is given by a more complicated expression 
\begin{equation} \label{78}
	<S^\mu_T> = \frac{\sqrt{m^2 + M^2}}{M} \left[ \left(1 - \frac{1}{2} \frac{m^2}{m^2+M^2} \right) v^\mu - \frac{1}{2} \frac{m^2}{m^2+M^2} \left( \dot{X}^\mu - \frac{1}{2} \frac{m^2}{m^2+M^2} \frac{n^\mu}{\dot{X}n} \right) \frac{v n}{\dot{X}n} \right],
\end{equation}
which is a consequence of the fact that $v^\mu$ no longer satisfies $\dot{X} v= 0$ but 
\begin{equation}
	\dot{X}v - \frac{1}{2} \frac{m^2}{\sqrt{m^2+M^2}} \frac{n v}{\dot{X}n} = 0.
\end{equation}
Notice that we still have $\dot{X} <S_T> = 0$. In the limit $m^2 << M^2$ the Liouville equation now gives
\begin{eqnarray}\label{80}
	\dot{v}^\mu &=& \frac{1}{M} \left( 1 - \frac{1}{2} \frac{m^2}{M^2} \right) (q + 2\mu) F^{\mu\nu} v_\nu + 2\mu \left[ \left(1 + \frac{1}{2} \frac{m^2}{M^2}  \right) \dot{X}^\mu - \frac{1}{2} \frac{m^2}{M^2} \frac{n^\mu}{\dot{X}n} \right] Fv\dot{X} + \nonumber \\
	&+& \frac{m^2}{M^2} \frac{q}{2M} F^{\mu\nu}n_\nu \frac{vn}{\dot{X}n} - \frac{1}{2M} \frac{m^2}{M^2} \left( 2\mu M \dot{X}^\mu + q \frac{n^\mu}{\dot{X}n} \right) \frac{Fvn}{\dot{X}n}.
\end{eqnarray}
We then take the time derivative in (\ref{78}) and use (\ref{80}) to find that $<\dot{S}^\mu_T>$ again obeys (\ref{68}). 

As a last remark we want to mention that the distribution function is also required to satisfy some sort of positivity condition \cite{Berezin:1976eg} like 
\begin{equation}
	\int d\Psi_5 d\Psi_3 d\Psi_2 d\Psi_1 d\Psi_0 \,\, \rho \,\, F^\star F \ge 0, 
\end{equation}
for any phase space function $F$. Like in the classical relativistic case \cite{Berezin:1976eg} our distribution functions do not satisfy a positivity condition. It seems that this can only be implemented when the spinning particle has internal degrees of freedom \cite{Barducci:1980xk}.

\section{Conclusions}\label{s6}

We discussed the inclusion of VSR like terms in a Lorentz invariant theory starting with the spinning particle model for a fermion. It provides a way to generate a class of Lorentz violating theories which have a preferred direction in space but at the same time keeps many essential elements of special relativity. Its effects appear at a scale $m$ where the anisotropy becomes relevant. Many terms invariant by VSR can be added to relativistic invariant equations and we developed a systematic way to generate such terms. In particular we determined how the BMT equation, which describes the electron spin precession in a electromagnetic field, is modified by VSR. We showed that in the rest frame the spin still precesses but VSR effects will now produce new effects. It has been argued that VSR is not consistent with Thomas precession \cite{Das:2009fi} but our analysis does not support this view.   It is well known that for a particle with $g=2$ in a magnetic field the spin precesses in such a way that the longitudinal polarization is constant, while the presence of a electric field in the relativistic limit makes the spin to precess very slowly. It would be interesting to find how VSR changes these properties.

\section{Acknowledgments}

The work of J.A. was partially supported by Fondecyt \# 1110378 and Anillo ACT 1102. He also wants to thank the Instituto de F\'{i}sica, USP and the IFT/SAIFR for its kind hospitality during his visits to S\~ao Paulo. The work of V.O.R. is supported by CNPq grant 304116/2010-6 and FAPESP grant 2008/05343-5. He also wants to thank Facultad de Fisica, PUC Chile for its kind hospitality during his visits to Santiago.


\begin{thebibliography}{999}


\bibitem{Colladay:1998fq}
D.~Colladay and V.~A. Kostelecky, {\it {Lorentz violating extension of the
  standard model}},  {\em Phys.Rev.} {\bf D58} (1998) 116002,
  [\href{http://xxx.lanl.gov/abs/hep-ph/9809521}{{\tt hep-ph/9809521}}].

\bibitem{Liberati:2013xla}
S.~Liberati, {\it {Tests of Lorentz invariance: a 2013 update}},
  \href{http://xxx.lanl.gov/abs/1304.5795}{{\tt arXiv:1304.5795}}.

\bibitem{Cohen:2006ky}
A.~G. Cohen and S.~L. Glashow, {\it {Very special relativity}},  {\em
  Phys.Rev.Lett.} {\bf 97} (2006) 021601,
  [\href{http://xxx.lanl.gov/abs/hep-ph/0601236}{{\tt hep-ph/0601236}}].

\bibitem{Cohen:2006ir} 
  A.~G.~Cohen and S.~L.~Glashow, {\it {A Lorentz-Violating Origin of Neutrino Mass?}}
[\href{http://xxx.lanl.gov/abs/hep-ph/0605036}{{\tt hep-ph/0605036}}].

\bibitem{Cohen:2006sc}
A.~G. Cohen and D.~Z. Freedman, {\it {SIM(2) and SUSY}},  {\em JHEP} {\bf 0707}
  (2007) 039, [\href{http://xxx.lanl.gov/abs/hep-th/0605172}{{\tt
  hep-th/0605172}}].

\bibitem{Vohanka:2011aa}
J.~Vohanka, {\it {Gauge Theory and SIM(2) Superspace}},  {\em Phys.Rev.} {\bf
  D85} (2012) 105009, [\href{http://xxx.lanl.gov/abs/1112.1797}{{\tt
  arXiv:1112.1797}}].

\bibitem{Gibbons:2007iu}
G.~Gibbons, J.~Gomis, and C.~Pope, {\it {General very special relativity is
  Finsler geometry}},  {\em Phys.Rev.} {\bf D76} (2007) 081701,
  [\href{http://xxx.lanl.gov/abs/0707.2174}{{\tt arXiv:0707.2174}}].

\bibitem{Muck:2008bd}
W.~Muck, {\it {Very Special Relativity in Curved Space-Times}},  {\em
  Phys.Lett.} {\bf B670} (2008) 95--98,
  [\href{http://xxx.lanl.gov/abs/0806.0737}{{\tt arXiv:0806.0737}}].

\bibitem{SheikhJabbari:2008nc}
M.~Sheikh-Jabbari and A.~Tureanu, {\it {Realization of Cohen-Glashow Very
  Special Relativity on Noncommutative Space-Time}},  {\em Phys.Rev.Lett.} {\bf
  101} (2008) 261601, [\href{http://xxx.lanl.gov/abs/0806.3699}{{\tt
  arXiv:0806.3699}}].

\bibitem{Das:2010cn}
S.~Das, S.~Ghosh, and S.~Mignemi, {\it {Noncommutative Spacetime in Very
  Special Relativity}},  {\em Phys.Lett.} {\bf A375} (2011) 3237--3242,
  [\href{http://xxx.lanl.gov/abs/1004.5356}{{\tt arXiv:1004.5356}}].

\bibitem{Ahluwalia:2010zn}
D.~Ahluwalia and S.~Horvath, {\it {Very special relativity as relativity of
  dark matter: The Elko connection}},  {\em JHEP} {\bf 1011} (2010) 078,
  [\href{http://xxx.lanl.gov/abs/1008.0436}{{\tt arXiv:1008.0436}}].

\bibitem{Chang:2013xwa}
Z.~Chang, M.-H. Li, X.~Li, and S.~Wang, {\it {Cosmological model with local
  symmetry of very special relativity and constraints on it from supernovae}},
  \href{http://xxx.lanl.gov/abs/1303.1593}{{\tt arXiv:1303.1593}}.

\bibitem{Cheon:2009zx}
S.~Cheon, C.~Lee, and S.~J. Lee, {\it {SIM(2)-invariant Modifications of
  Electrodynamic Theory}},  {\em Phys.Lett.} {\bf B679} (2009) 73--76,
  [\href{http://xxx.lanl.gov/abs/0904.2065}{{\tt arXiv:0904.2065}}].


\bibitem{Bargmann:1959gz}
V.~Bargmann, L.~Michel, and V.~Telegdi, {\it {Precession of the polarization of
  particles moving in a homogeneous electromagnetic field}},  {\em
  Phys.Rev.Lett.} {\bf 2} (1959) 435.

\bibitem{Jackson}
J.~Jackson, {\em {Classical Electrodynamics}}.
\newblock {Willey, New York}, {2nd}~ed., {1975}, Ch. 11.

\bibitem{Berezin:1976eg} 
  F.~A.~Berezin and M.~S.~Marinov,
  ``Particle Spin Dynamics as the Grassmann Variant of Classical Mechanics,''
  Annals Phys.\  {\bf 104}, 336 (1977).

\bibitem{Howe:1988ft}
P.~S. Howe, S.~Penati, M.~Pernici, and P.~K. Townsend, {\it {Wave Equations for
  arbitrary spin from quantization of the extended supersymmetric spinning
  particle}},  {\em Phys.Lett.} {\bf B215} (1988) 555.

\bibitem{Pierri:1990rp}
M.~Pierri and V.~O. Rivelles, {\it {BRST Quantization of Spinning Relativistic
  Particles with Extended Supersymmetries}},  {\em Phys.Lett.} {\bf B251}
  (1990) 421--426.

\bibitem{Alfaro:2013uva}
J.~Alfaro and V.~O. Rivelles, {\it {Non Abelian Fields in Very Special
  Relativity}}, Phys.\ Rev.\ D {\bf 88}, 085023 (2013)
  [arXiv:1305.1577 [hep-th]].
	
	
\bibitem{Barducci:1982yw}
A.~Barducci, {\it {Pseudoclassical description of relativisitic spinning
  particles with anomalous magnetic moment}},  {\em Phys.Lett.} {\bf B118}
  (1982) 112.

\bibitem{Gitman:1992an}
D.~Gitman and A.~Saa, {\it {Quantization of spinning particle with anomalous
  magnetic momentum}},  {\em Class.Quant.Grav.} {\bf 10} (1993) 1447--1460,
  [\href{http://xxx.lanl.gov/abs/hep-th/9209086}{{\tt hep-th/9209086}}].

\bibitem{Deriglazov:2011gy}
A.~Deriglazov, {\it {Semiclassical Description of Relativistic Spin without use
  of Grassmann variables and the Dirac equation}},  {\em Annals Phys.} {\bf
  327} (2012) 398--406, [\href{http://xxx.lanl.gov/abs/1107.0273}{{\tt
  arXiv:1107.0273}}].

\bibitem{Das:2009fi}
S.~Das and S.~Mohanty, {\it {Very Special Relativity is incompatible with
  Thomas precession}},  {\em Mod.Phys.Lett.} {\bf A26} (2011) 139--150,
  [\href{http://xxx.lanl.gov/abs/0902.4549}{{\tt arXiv:0902.4549}}].

\bibitem{Barducci:1980xk} 
  A.~Barducci, R.~Casalbuoni and L.~Lusanna,
  ``Anticommuting Variables, Internal Degrees of Freedom, and the Wilson Loop,''
  Nucl.\ Phys.\ B {\bf 180}, 141 (1981).

\end{thebibliography}
\end{document}